\documentclass[aps,amsmath,prl,reprint,showpacs,preprintnumbers,twocolumn]{revtex4}
\usepackage{amsfonts}
\usepackage{graphicx}
\usepackage{dcolumn}
 \usepackage{pdfsync}
\usepackage{multirow}
\usepackage{subfigure}
\usepackage{palatino}
\usepackage{graphicx}
\usepackage{epsfig}
\usepackage{ifpdf}
\usepackage{color}
\ifpdf \DeclareGraphicsExtensions{.pdf,.png,.jpg,.mps}
 \else
\DeclareGraphicsExtensions{.eps,.pdf} \fi

\def\be{\begin{eqnarray}}
\def\ee{\end{eqnarray}}

\def\Xe#1{$^{#1}$Xe}

\def\Thad#1{{#1}}

\begin{document}

\title{A Laboratory Search for a Long-Range T-odd, P-odd Interaction from Axion-Like Particles using Dual Species Nuclear Magnetic Resonance with Polarized $^{129}$Xe and $^{131}$Xe Gas}

\author{M. Bulatowicz}
\author{R. Griffith}
\author{M. Larsen} 
\author{J. Mirijanian}
\affiliation{Northrop Grumman Corporation, Woodland Hills, California 91367, USA} 
\author{C.B. Fu} 
\author{E. Smith}
\author{W. M. Snow} 
\author{H. Yan}
\affiliation{Indiana University, Bloomington, Indiana 47408, USA}
\affiliation{Center for Exploration of Energy and Matter, Indiana University, Bloomington, IN 47408}
\author{T. G. Walker}
\affiliation{University of Wisconsin, Madison, Wisconsin 53706, USA}

\begin{abstract}

Various theories beyond the Standard Model predict new particles with masses in the sub-eV range with very weak couplings to ordinary matter. A new $P$-odd and $T$-odd  interaction between polarized and unpolarized nucleons proportional to ${\vec{K}} \cdot {\vec{r}}$ is one such possibility, where ${\vec{r}}$ is the distance between the nucleons and ${\vec{K}}$ is the spin of the polarized nucleon. Such an interaction involving a scalar coupling $g_{s}$ at one vertex and a pseudoscalar coupling $g_{p}$ at the polarized nucleon vertex can be induced by the exchange of spin-$0$ bosons. We used the NMR cell test station at Northrop Grumman Corporation to search for NMR frequency shifts in polarized $^{129}$Xe and $^{131}$Xe when  a non-magnetic zirconia rod is moved near the NMR cell.  Long ($T_{2}\sim20$ s)  spin-relaxation times allow precision measurements of the NMR frequency ratios, which are insensitive to magnetic field fluctuations. Ê Combined with existing theoretical calculations of the neutron spin contribution to the nuclear angular momentum in xenon nuclei,Êthe measurements improve the laboratory upper bound on the product $g_sg_p^n$ by  \Thad{two orders of magnitude} for distances near 1 mm.  \Thad{The sensitivity of this technique can be increased by at least two more orders of magnitude.}

\end{abstract}


\maketitle


The possible existence of new interactions of nature with ranges of macroscopic scale (millimeters to microns) and very weak couplings to matter was suggested long ago~\cite{Leitner,Hill} and has recently begun to attract more scientific attention. Particles which might transmit such interactions are starting to be \Thad{referred to generically as WISPs (Weakly-Interacting sub-eV Particles)~\cite{Jae10}}. Many theories beyond the Standard Model possess extended symmetries which, when broken at a high energy scale, produce pseudo-Nambu-Goldstone bosons and lead to weakly-coupled light particles with relatively long-range interactions  such as axions, familons, and Majorons~\cite{PDG12}.  Several attempts to explain dark matter and dark energy also produce new weakly-coupled long-range interactions.  The fact that the dark energy density of $O$(1 meV$^4$)  corresponds to a length scale of $~100$ $\mu$m  encourages searches for new phenomena around this scale~\cite{Ade09, Antoniadis11}.

Long ago Moody and Wilczek~\cite{Moody84} considered the form of interactions which could be induced by the exchange of a spin $0$ field between fermions with scalar or pseudoscalar couplings.  They highlighted an interesting  $P$ and $T$ violating scalar-pseudoscalar (monopole-dipole) interaction of the form
\begin{equation}\label{eq.potential.org}
V=\hbar^{2}g_{s}g_{p}\frac{\hat{\mathbf{\sigma}}\cdot\mathbf{\hat{r}}}{8\pi m_n}
\left(
\frac{1}{r\lambda}+\frac{1}{r^{2}}
\right)
e^{-r/\lambda},
\end{equation}
where $m$ is \Thad{ the  mass  and  ${\bf K}=\hbar\hat{\sigma}/2$ is the spin} of the polarized particle, $\lambda$ is the interaction range, $\hat{\mathbf{r}}={\mathbf{r}}/r$ is the unit vector between the particles, and $g_{s}$ and $g_{p}$ are the scalar and pseudoscalar coupling constants. Axions \cite{Pec77},  axion-like particles~\cite{Jae10} predicted by string theory~\cite{Svrcek2006}, or a very light spin-1 boson~\cite{Fayet96}, can induce such an interaction and are candidates for cold dark matter~\cite{Kolb1990}. The axion mass is constrained to the so-called ``axion window"~\cite{Antoniadis11} between 1 $\mu$eV to 1 meV, corresponding to a range between 2 cm to 20 $\mu$m. Most experiments that have  searched for such interactions~\cite{Ni99,You96,Vas09,Gle08, Ham07, Rit93} are sensitive to ranges $\lambda \geq 1$ cm. 

Recent experimental work to constrain the monopole-dipole interaction at shorter distances has employed ultracold neutron bound states~\cite{Bae07, Jenke12}, polarized ultracold neutrons in material traps~\cite{Ser09, Ig09}, and spin relaxation of polarized nuclei~\cite{Pok10, Pet10, Fu11, Zhe12}, \Thad{with the most stringent direct experimental constraint on the  product $g_sg_p^n$ involving the neutron pseudoscalar coupling $g_p^n$ coming from NMR measurements on polarized $^{3}$He~\cite{Pet10,Chu}}. Torsion balance measurements  recently set new stringent limits on possible monopole-dipole interactions involving polarized electrons~\cite{Hoedl11}.  Although all of these laboratory results are less stringent than those inferred from separate limits on $g_{s}$ and $g_{p}^n$ through a combination of torsion balance experiments  and astrophysical constraints from stellar evolution and SN1987A~\cite{Raffelt12}, laboratory experiments are more direct and therefore are of fundamental interest.  

We used an NMR cell test station at Northrop-Grumman Corp.,  developed to evaluate dual species xenon cells for a gyroscope, to search for NMR frequency shifts in \Thad{polarized $^{129}$Xe  and $^{131}$Xe from the monopole-dipole interaction} when a  zirconia rod is moved near the NMR cell. By comparing the simultaneous frequencies of the two Xe isotopes, magnetic field changes are distinguished from frequency shifts due to the monopole-dipole coupling of the polarized Xe nuclei to the zirconia rod.  We use  calculations~\cite{Ressel97,Menendez12}  of the neutron spin contribution to the nuclear angular momentum in $^{129}$Xe  and $^{131}$Xe
 to put a new upper bound on the product  $g_{s}g_{p}^{n}$ for ranges at the mm scale. 

\begin{figure}{}
\begin{center}
\includegraphics[width=3.4 in]{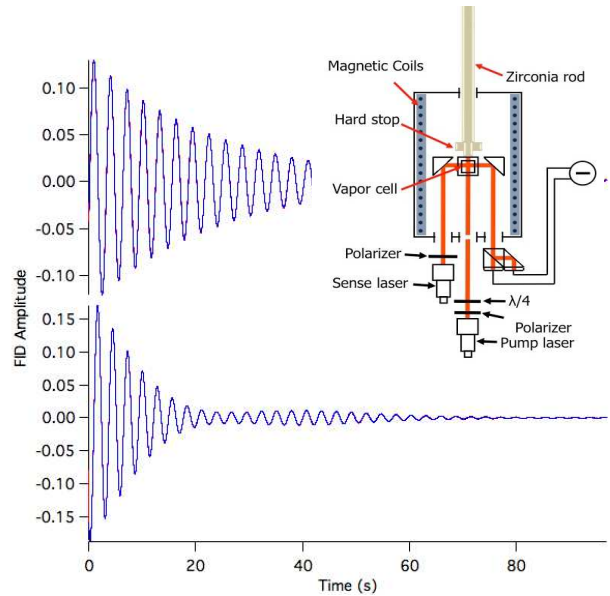}
	\caption{Dual species free-induction-decay (FID) spectrometer. A 2 mm glass cell containing $^{129}$Xe and $^{131}$Xe polarized by spin-exchange optical pumping of $^{85}$ Rb is in a uniform magnetic field inside a magnetic shield. A moveable zirconia rod enters from one end of the apparatus and the pump and sense lasers enter from the opposite side. FID signals of the  Xe isotopes are sensed optically using the Rb vapor as a magnetometer.  Sample FIDs, with fits, show $^{129}$Xe above, $^{131}$Xe below. For display purposes, the FID oscillations are down-converted to about 0.3 Hz from 152 or 45 Hz.}
\label{fig:Apparatusconcept}
\end{center}
\end{figure}

The NMR measurements use a {magnetically shielded} 
\Thad{Pyrex} co-magnetometer cell consisting of $^{85}$Rb, 5 Torr  $^{129}$Xe, 45 Torr  $^{131}$Xe, and 250 Torr N$_2$. 
{Fig.~\ref{fig:Apparatusconcept} depicts the apparatus. The Rb atoms are spin-polarized by absorption of circularly polarized light from a \Thad{795 nm, 50 mW pump diode }laser and, through spin-exchange collisions,  polarize the Xe nuclei parallel to a DC magnetic field \cite{WalkerRMP}.  
The Rb atoms also serve as a magnetometer 
that  detects the precession of the two xenon isotopes; the transverse magnetic fields of the polarized nuclei produce an oscillating transverse spin polarization of the Rb atoms that is detected as a rotation of the polarization  of the linearly polarized sense laser. }

We record simultaneous free-induction-decay (FID) signals from the Xe nuclei (Fig. 1). We initiate the  FIDs by applying short resonant $\sim\pi/4$ pulses  to rotate the two Xe polarizations away from the $\hat{z}$-direction.  
\Thad{ We isolate a single species FID by taking the raw FID, Fourier transforming it,  applying a 10 Hz band-pass filter centered at the resonance frequency (152 Hz for $^{129}$Xe, 45 Hz for $^{131}$Xe), and inverse-transforming it.}
 We eliminate the first $1.6$ seconds and the last $0.2$ seconds of this signal, which contain filter artifacts and pulse residuals.  Over the 100 sec FID acquisition time, there are Zeeman frequency drifts that \Thad{arise from alkali polarization drifts and actual magnetic field noise.  We use the $^{129}$Xe signal as our magnetometer for isolating these effects,   fitting the $^{129}$Xe FIDs to the form $S_{1}=A\cos(\alpha(t) +\phi_1)e^{-t/T_{2}}$ where $\alpha(t)=2\pi |\gamma_1|\int_0^t B(t')dt'$, and $\gamma_1$ is the gyromagnetic ratio. We use  the deduced magnetic field $B(t)$ as input for the fits to the $^{131}$Xe FIDs to infer the $^{131}$Xe frequency $f_2$ as described in more detail below.}

 We performed the $g_s g_p^n$ search by comparing the ratios $f_1/f_2$ when a mass was close to or far away from the cell.  \Thad{To this end
we attached a mounting structure to the magnetic shielding ( DC shielding factor $\approx 3 \times 10^{4}$) to allow the repeatable positioning of a zirconia rod  either $750 \pm 60$ microns, or $1.0 \pm 0.$1 cm, from the inner surface of the cell.} The  cell was a $1.9$ mm internal dimension cube with a $0.5$ mm diameter fill stem pinch-off {on one side} 
 and  $0.55$ mm thick walls. 
{The rod position was alternated for successive runs, with 
the time from the start of one run to the start of the next being about two minutes ($4-5$ \Thad{ relaxation times}). Either the pump light polarization or the magnetic field direction was reversed every 10th run, and the magnitude of the magnetic field was changed once, from 0.13 to 0.064 G. }

The \Thad{$K=1/2$} $^{129}$Xe isotope couples only to vector fields such as the external magnetic field, whereas the 
\Thad{$K=3/2$} $^{131}$Xe isotope { also experiences quadrupole interactions with the electric field gradients on the cell walls.  The nuclear spin Hamiltonian for \Xe{129} ($j=1$) and  \Xe{131} ($j=2$) is
\be
H_j=h\gamma_jB_jK_z+hX_jK_z+\delta_{j,2}H_Q.
\ee
 There are $4$ NMR frequencies deduced from the FIDs: $f_1=|\gamma_{1}|B_{1}+\hat z\cdot \hat BX_{1}$, $f_{2i}=|\gamma_{2}|B_{2}-\hat z\cdot \hat BX_{2}+q_i$ where the gyromagnetic ratios of species $j$ are $\gamma_j$, the monopole-dipole couplings are $X_j$, and the quadrupole shifts of the 3 NMR lines of $^{131}$Xe are $q_i$ ($i=-1,0,1$).  The presence of the three frequencies is evident in the unusual shape of the $^{131}$Xe FID in Fig.~\ref{fig:Apparatusconcept}.  The  two Xe isotopes may experience slightly different magnetic fields $B_j$ {due to different interactions with the polarized Rb atoms}. In previous experimental work using NMR of $K=3/2$ nuclei \cite{Lamoreaux86, Majumder90, Venema92},  quadrupole interactions were  treated in a spherical approximation. { Our cell is  cubic but the $6$ internal surfaces are not identical, and the glass stem is located 
on one of the cell faces whose normal is along the y-direction.} The averaged quadrupole interaction on a surface with normal $\bf\hat{n}$ is $V_n={Q_n }\left({\bf K\cdot \hat{n}\hat{ n}\cdot K-K\cdot K}/3\right)/2$~\cite{Wu87}. We therefore consider the quadrupole interaction as a sum of contributions from each of the orthogonal faces of the cell.  2nd order perturbation theory shows that the three resonance frequencies in the \Xe{131} free-induction decay are
\def\abs#1{{\left|#1\right|}}
\be
f_{2i}=\abs{\gamma_2 B_2}-\hat z\cdot \hat BX_2+2i\abs{Q_{\|}}+{3Q_\perp^2{\delta_{i0}}\over 4f_2}
\ee 
{where $Q_\|=Q_z-(Q_x+Q_y)/2$ and  $Q_\perp=Q_x-Q_y$. The first and second order quadrupole shifts are typically 20 mHz and 500 $\mu$Hz respectively. By taking the average 
\be
f_2={f_{21}+f_{2-1}\over 2}=\abs{\gamma_2 B_2}-\hat z\cdot \hat BX_2
\ee
the quadrupole interaction is  cancelled to second order.}
\Thad{We therefore fit the $^{131}$Xe FIDs to 
\be
\!\!S_2\!\!&=&\!\!\sum_i \!\!A_i e^{-t/T_{2i}}\!\! \cos\!\left({f_2\over f_1}\alpha_1(t)+2\pi(f_{2i}-f_2)t+\phi_{2i}\right)
\ee
with the quadrupole shifts and the frequency ratio being the fit parameters of primary interest.
}

The magnetic field experienced by the two nuclei is not only the applied field, but contains an additional contribution $B_{Aj}P$, $B_{Aj}\approx 0.1$ mG, from the  alkali atoms of polarization $P$\cite{Schaefer89}.  To the extent that the alkali field is independent of the Xe isotope, the resulting noble gas frequency shift will be proportional to the nuclear gyromagnetic ratio \Thad{and  cancel in the frequency ratio. However, as} shown in Fig.~\ref{fig:isotope},  we observe a small  shift in the frequency ratio that is correlated with the product of the sign of the light polarization and the magnetic field direction. This signals the presence of a slight isotopic difference in the alkali fields experienced by the two nuclei, to our knowledge the first observation of this phenomenon. {We account for this   by the parameterization  $B_{A1}=B_A(1+\delta B_A/2)$, $B_{A2}=B_A(1-\delta B_A/2)$}.
{ To first order in $\delta B_A$ and $X$, the frequency ratio is 
\be
{f_1\over f_2}&=&\varrho\left(1+\delta B_A  {PB_A\over B_0}\right)+ \hat z\cdot \hat B{X_1+\varrho X_2\over f_2}\label{eq:fratio}
\ee 
where the ratio of gyromagnetic ratios is $\varrho={|\gamma_1|/|\gamma_2|}$.  A monopole-dipole coupling causes a rod in/rod out change in $f_1/f_2$ that reverses with magnetic field direction and is independent of polarization. }  The isotope effect, on the other hand, correlates with the product of the magnetic field direction and the polarization direction.  In Fig.~\ref{fig:isotope}, the magnetic field magnitude was halved at run 600, causing a doubling in the modulation amplitude that is reproduced in the data, consistent with Eq.~\ref{eq:fratio}.
The  fractional isotope shift in the alkali field, deduced only from runs with the rod out,  is $\delta B_A=0.0017(1)$. \Thad{An explanation for the isotope shift  notes that the alkali field includes contributions from RbXe molecules.  In these molecules different hyperfine interactions cause  a slightly different alkali field for the two Xe isotopes.}

\Thad{We also observe (Fig.~\ref{fig:isotope}) a small (1.5 ppm) change in the apparent ratio of gyromagnetic ratios when the magnitude of the applied magnetic field is halved.  Since the second order quadrupole shift is inversely proportional to magnetic field, the change in the ratio of gyromagnetic ratios may be a sign of imperfect removal of quadrupole shifts.
} 

\begin{figure}
\begin{center}
	\includegraphics[width=3.3 in]{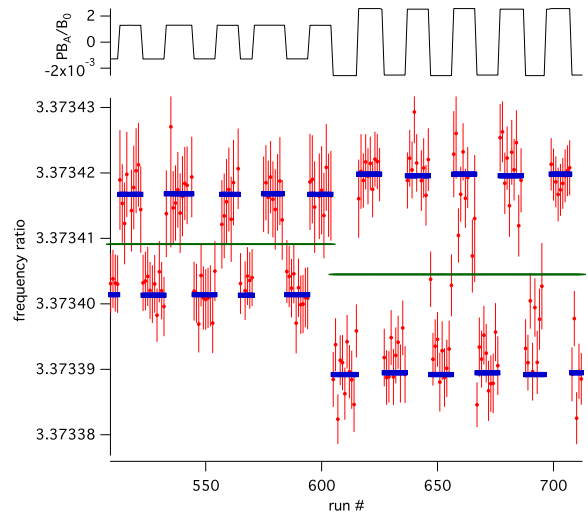}
	\caption{A sampling of frequency ratio measurements, showing a correlation with the product of the magnetic field direction and the polarization direction, as expected for a differential alkali field shift.   \Thad{The modulation doubles, as expected, when the magnetic field is halved at run 605, but a 1.5 ppm shift in the ratio of gyromagnetic ratios is also observed, as indicated by the solid green lines.}}
	\label{fig:isotope}
	\end{center}
\end{figure}

Figure \ref{fig:Frequencyratio}a shows the full sequence of frequency ratio measurements, corrected for the differential alkali field shift \Thad{and the observed shift when the magnitude of the magnetic field was halved}.  The data were taken over 4 days, with varying degrees of magnetic noise and optimization of magnetometer signals accounting for variations in the noise levels.  \Thad{About 70 runs were discarded due to rapid magnetic field variations that our fitting was not able to account for, or  rapidly varying alkali field shift that were probably due to laser instabilities, especially in the set of runs shortly after run 200.}  We fit the corrected frequency ratio data to the function $\varrho+\hat{z}\cdot \hat{B}X/f_2$ (Fig. \ref{fig:Frequencyratio}b) with  the ratio of gyromagnetic ratios $\varrho$ and the weighted  monopole-dipole shift $X=X_1+\varrho X_2$.  The \Thad{mean of the gyromagnetic ratios at the two field strengths is $\varrho=3.3734072(5)$, and
$X=1(7)$ $\mu{\rm Hz}$.}
The value of $\varrho$ is consistent with  Brinkmann \cite{Brinkmann}, who found $3.37340(4)$ and observed a pressure dependence. We measured only one cell and have no information about pressure effects, so our value of $\varrho$  should be considered specific to the  parameters of our cell. It may also be affected by an imperfect model for the quadrupole interaction. 


\begin{figure}
\begin{center}
	\includegraphics[width=3.0 in]{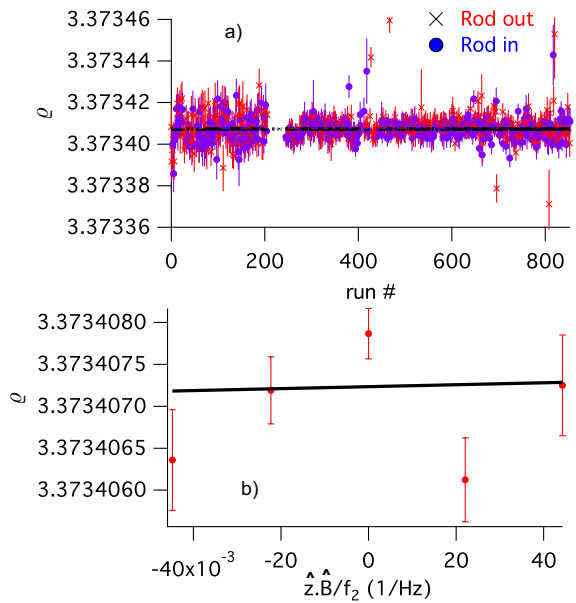}
	\caption{a) Frequency ratio measurements obtained with a variety of magnetic field and polarization directions, and two magnetic field strengths, with and without a nearby zirconia rod. b) Average frequency ratios for different magnetic field directions and magnitudes.  The slope of the solid line,  the monopole-dipole frequency shift, is consistent with zero.}
	\label{fig:Frequencyratio}
	\end{center}
\end{figure}

To  use $X$ to set a limit on the monopole-dipole interaction, we write Eq.~\ref{eq.potential.org} as $V=v({\bf r})\langle \sigma_z\rangle$, giving $hX_i=\langle v({\bf r})\rangle_{\rm vol}\langle \sigma_z\rangle_i/K_i$.  The volume average $\langle v({\bf r})\rangle_{\rm vol}$ over the polarized xenon and unpolarized zirconia (nucleon density  $3.6 \times10^{24}\ {\rm cm}^{-3}$) mass distributions  is done numerically.  The 
spin expectation values require knowledge of the neutron and proton angular momentum distributions of $^{129}$Xe and $^{131}$Xe}. Such calculations \cite{Menendez12,Engel91, Ressel97, Toivanen09} have been done  in the context of experimental searches for dark matter spin flip scattering in liquid xenon. All calculations confirm that the spin of these even-Z nuclei almost entirely comes from the neutrons, so we interpret our results in terms of neutron coupling.  In particular, the Ressel and Dean \cite{Ressel97} Bonn A calculations, in good agreement with recent effective field theory analysis of Menendez et al.~\cite{Menendez12}, give {$\langle K_{n}^{129}\rangle=0.359$, $\langle K_{p}^{129}\rangle=0.028$, $\langle K_{n}^{131}\rangle=-0.227$, $\langle K_{p}^{131}\rangle=-0.009$, and $\varrho=3.065$. These numbers give $hX=0.51\langle v({\bf r})\rangle_{\rm vol}$.

Figure~\ref{fig:gsgpconstraints}  shows how our upper limit on $X$ results in a limit on the monopole-dipole coupling strength as a function of the range $\lambda$.  It excludes the light grey area in the ($g_{s}g_{p}^n$, $\lambda$) plane.  The limits shown involve the pseudoscalar coupling $g_{p}^n$ of the neutron and the scalar coupling $g_{s}$ of an ensemble of unpolarized matter made of a roughly equal proportion of \Thad {neutrons, protons, and electrons}.   \Thad{Our bounds $g_sg_p^n \leq 10^{-18}$ for  $\lambda=0.1$ mm to  $g_sg_p^{n} \leq 2 \times 10^{-24}$ }for $\lambda=1$ cm
 are the most stringent laboratory limit 
over these distances, which correspond to exchange boson masses from $10^{-3}$ to $10^{-5}$ eV/$c^{2}$.

\begin{figure}{}
\begin{center}
		\includegraphics[width=3.3 in]{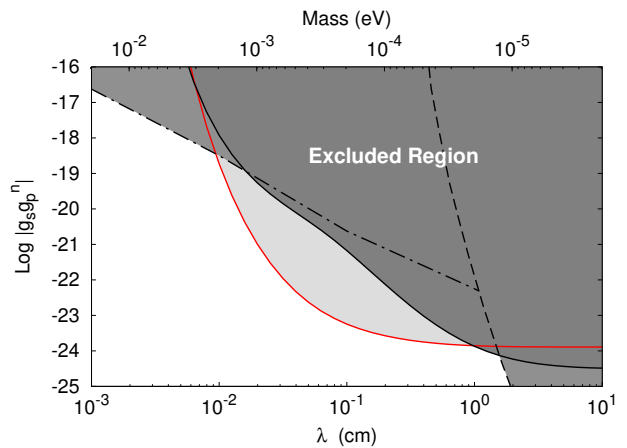}
	\caption{Constraints on the coupling constants $g_sg_p^n$ of the spin-dependent force as a function of the force range $\lambda$. The dashed line is  from~\cite{You96}, the dash-dotted line is from~\cite{Pet10},  the solid line is from~\cite{Chu}, and the red solid line is the present work. }
	\label{fig:gsgpconstraints}
	\end{center}
\end{figure}

There is great room for improvement of the measurement sensitivity using this technique.  Direct stabilization of the $^{129}$Xe frequency using magnetic field feedback, plus stabilization of the  alkali magnetization, would \Thad{likely reduce the statistical uncertainties by 1-2 orders of magnitude and lower the limits on $g_sg_p^n$ correspondingly.  Improvement of the magnetic field shielding and making NMR oscillators would yield additional sensitivity improvements.}

\Thad{{\it Note added}:  Since the submission of this paper, a preprint \cite{Tullney2013} reports improved limits on $g_sg_p^n$ at distances longer than 0.5 mm.  }

 

This work is supported by  NGC IRAD funding, the Department of Energy, and NSF grants PHY-0116146 and PHY-1068712.  C. B. Fu, E. Smith, W. M. Snow, and H. Yan acknowledge support from the Indiana University Center for Spacetime Symmetries.  We acknowledge helpful suggestions from the referees.


\begin{thebibliography}{99}

\def\NIM{\rm Nucl. Instr. Meth.}
\def\NIMA{{\rm Nucl. Instr. and Meth.} A} 
\def\NPB{{\rm Nucl. Phys.} B} 
\def\PLB{{\rm Phys. Lett.}  B} 
\def\PRL{\rm Phys. Rev. Lett.}
\def\PL{\rm Phys. Lett.}
\def\PLA{\rm Phys. Lett. A} 
\def\PRD{{\rm Phys. Rev.} D} 
\def\PRA{{\rm Phys. Rev.} A}
\def\PRC{{\rm Phys. Rev.} C} 
\def\PR{\rm Phys. Rev.}
\def\NPB{{\rm Nucl. Phys.} B}
\def\MRI{\rm J. Magn. Res. Imag.}
\def\MRM{\rm Magn. Res. Med.}
\def\PRT{\rm Phys. Rep.}
\def\LTP{\rm J. Low Temp. Phys.}
\def\EPJD{\rm Eur. Phys. J. D}
\def\EPJA{\rm Eur. Phys. J. A}
\def\EPL{\rm Europhys. Lett.}
\def\MRM{\rm Mag. Res. Med.}
\def\JPF{\rm J. Phys. III France}
\def\JMR{\rm J. Magn. Res.}
\def\JMRB{\rm J. Magn. Res. B}
\def\MRM{\rm Magn. Res. Med.}
\def\Journal#1#2#3#4{{#1} {\bf #2}, #3 (#4)}

\bibitem{Leitner}J. Leitner and S. Okubo, \Journal{\PR}{136}{B1542}{1964}.
\bibitem{Hill}C.T. Hill and G. G. Ross, \Journal{\NPB}{311}{253}{1988}.
\bibitem{Jae10}  J. Jaeckel and A. Ringwald, Annu. Rev. Nucl. Part. Sci. {\bf 60}, 405 (2010). 
\bibitem{PDG12}K. Nakamura {\it et al.} (Particle Data Group), J. Phys. G {\bf37}, 075021 (2010) and partial update for the 2012 edition. 
\bibitem{Ade09}  E. G. Adelberger et al., Prog. Part. Nucl. Phys. {\bf 62}, 102 (2009). 
\bibitem{Antoniadis11} I. Antoniadis {\it et al.}, C. R. Physique {\bf 12}, 755 (2011).
\bibitem{Moody84} J. E. Moody and F. Wilczek, Phys. Rev. D {\bf 30}, 130 (1984). 
\bibitem{Pec77} R. D. Peccei and H. R. Quinn, Phys. Rev. Lett. {\bf 38}, 1440 
(1977). 
\bibitem{Svrcek2006} P. Svrcek and E. Witten, J. High Energy Phys. {\bf 06}, 051 (2006).
\bibitem{Fayet96} P. Fayet, Class. Quant. Gravit. A {\bf 13}, 19 (1996).
\bibitem{Kolb1990} E.W. Kolb and M.S. Turner, {\it The Early Universe} (Addison-Wesley), Redwood, CA, (1990).
\bibitem{Ni99} W.-T. Ni, S.-S. Pan, H.-C. Yeh, L.-S. Hou, and J. Wan, Phys. Rev. Lett. {\bf 82}, 2439 (1999). 
\bibitem{You96} A. N. Youdin, D. Krause, K. Jagannathan, L. R. Hunter, S. K. Lamoreaux,  Phys. Rev. Lett. {\bf 77}, 2170 (1996).
\bibitem{Vas09} G.~Vasilakis, J.~M.~Brown, T.~W.~Kornack, and M.~V.~Romalis, Phys. Rev. Lett. {\bf 103}, 261801 (2009).
\bibitem{Gle08} A.~G.~Glenday, C.~E.~Cramer, D.~F.~Phillips, and R.~L.~Walsworth, Phys. Rev. Lett. {\bf 101}, 261801 (2008).
\bibitem{Ham07} G. D. Hammond, C. C. Speake, C. Trenkel, and A. P. Paton, Phys. Rev. Lett. {\bf 98}, 081101 (2007). 
\bibitem{Rit93} R. C. Ritter, L. I. Winkler, and G. T. Gillies, Phys. Rev. Lett. {\bf 70}, 701 (1993).
\bibitem{Bae07} S. Baessler, V. V. Nesvizhevsky, K. V. Protasov, and A. Y. Voronin, Phys. Rev. D {\bf 75}, 075006 (2007).
\bibitem{Jenke12} T. Jenke {\it et al.}, arxiv: 1208.3857v1 [hep-ex] (2012). 
\bibitem{Ser09} A. Serebrov, Physics Letters B {\bf 680}, 423 (2009).
\bibitem{Ig09} V. K. Ignatovich and Y. N. Pokotilovski, Eur. Phys. J. C {\bf 64}, 19 (2009).
\bibitem{Pok10} Y. N. Pokotilovski, Phys. Lett. B {\bf 686}, 114 (2010).
\bibitem{Fu11} C. Fu, T. R. Gentile, and W. M. Snow, Phys. Rev. D{\bf 83}, 031504(R) (2011).
\bibitem{Zhe12} W. Zheng, H. Gao, B. Lalremruata, Y. Zhang, G. Laskaris, C.B. Fu, and W.M. Snow, Phys. Rev. D {\bf 85}, 031505(R) (2012).
\bibitem{Pet10} A. K. Petukhov, G. Pignol, D. Jullien, and K. H. Andersen, Phys. Rev. Lett. {\bf 105}, 170401 (2010).
\bibitem{Chu} P. Chu {\it et al.}, \Thad{ Phys. Rev. D {\bf 87}, 011105(R) (2013).}
\bibitem{Hoedl11} S. A. Hoedl, F. Fleischer, E.G. Adelberger, and B. R. Heckel, Phys. Rev. Lett. {\bf 106}, 041801 (2011).
\bibitem{Raffelt12} G. Raffelt, Phys. Rev. D {\bf 86}, 015001 (2012).
\bibitem{Ressel97} M. T. Ressel and D. J.  Dean, Phys. Rev. C {\bf 56}, 535 (1997).
\bibitem{Menendez12} J. Menendez, D. Gazit, and A. Schwenk, \Thad{ Phys. Rev. D {\bf 86}, 103511 (2012).}
\bibitem{WalkerRMP} {T. G. Walker and W. Happer, Rev. Mod. Phys. {\bf 69}, 629 (1997).}
\bibitem{Lamoreaux86} S. K. Lamoreaux, J. P. Jacobs, B. R. Heckel, F. J. Raab, and E. N. Fortson, Phys. Rev. Lett. {\bf 57}, 3125 (1986). 
\bibitem{Majumder90} P. K. Majumder,  B. J. Venema, S. K. Lamoreaux, B. R. Heckel, and E. N. Fortson, Phys. Rev. Lett. {\bf 65}, 2931 (1990).
\bibitem{Venema92} B. J. Venema, P. K. Majumder, S. K. Lamoreaux, B. R. Heckel, and E. N. Fortson, Phys. Rev. Lett. {\bf 68}, 135 (1992).
\bibitem{Wu87} Z. Wu, W. Happer, and J. M. Daniels, Phys. Rev. Lett., {\bf 59}, 1480 (1987).
\bibitem{Schaefer89} {S. R. Schaefer, G. D. Cates, T-R. Chien, D. Gonatas, W. Happer,  and T. G. Walker, Phys. Rev. A 39, 5613 (1989).}
\bibitem{Brinkmann}D. Brinkmann,  Helv. Phys. Acta {\bf  36}, 413 (1963).
\bibitem{Engel91} J. Engel, Phys. Lett. {\bf 264}, 114  (1991).
\bibitem{Toivanen09} P. Toivanen, M. Kortelainen, J. Suhonen, and J. Toivanen, Phys. Rev. C {\bf 79}, 044302 (2009).
\bibitem{Tullney2013} K. Tullney, F. Allmendinger, M. Burghoff, W. Heil, S. Karpuk, W. Kilian, S. Knappe-GrŸneberg, W. MŸller, U. Schmidt, A. Schnabel, F. Seifert, Y. Sobolev, and L. Trahms, ArXiv:1303.6612.





\end{thebibliography}
\end{document}